\documentclass[twocolumn,superscriptaddress, preprintnumbers,pra]{revtex4}
\usepackage{graphicx}
\usepackage{dcolumn}
\usepackage{amssymb}
\usepackage{amsmath}
\usepackage{bm}
\usepackage{setspace}

\usepackage{hyperref}

\hypersetup
{
colorlinks=true, 
citecolor=blue
}

\begin{document}

\title{Dynamics of quantum Hall interfaces}

\author{Qi Li}
\affiliation{GBA Branch of Aerospace Information Research Institute, Chinese Academy of Sciences, Guangzhou 510535, China}

\author{Ken K. W. Ma}
\email[Corresponding author: ]{ken\textunderscore ma@magnet.fsu.edu}
\affiliation{National High Magnetic Field Laboratory and Department of Physics,
Florida State University, Tallahassee, Florida 32306, USA}

\author{Ruojun Wang}
\affiliation{National High Magnetic Field Laboratory and Department of Physics,
Florida State University, Tallahassee, Florida 32306, USA}

\author{Zi-Xiang Hu}
\affiliation{Department of Physics, Chongqing University,Chongqing 401331, People’s Republic of China}
\affiliation{Chongqing Key Laboratory for Strongly Coupled Physics, Chongqing 401331, People’s Republic of China }

\author{Hao Wang}
\affiliation{Shenzhen Institute for Quantum Science and Engineering, Southern University of Science and Technology, Shenzhen 518055, China}

\author{Kun Yang}
\affiliation{National High Magnetic Field Laboratory and Department of Physics,
Florida State University, Tallahassee, Florida 32306, USA}


\begin{abstract}
A quantum Hall (QH) interface is different from an ordinary QH edge, as the latter has its location determined by the confining potential, while the former can be unpinned and behave like a free string.
In this paper, we demonstrate this difference by studying three different interfaces formed by (i) the Laughlin state and the vacuum, (ii) the Pfaffian state and the vacuum, and (iii) the Pfaffian and the anti-Pfaffian states. We find that string-like interfaces propagating freely in the QH system lead to very different dynamical properties from edges. This qualitative difference gives rise to fascinating physics and suggests a different direction in future research on QH physics. We also discuss briefly possible analogies between QH interfaces and concepts in string theory.
\end{abstract}

\date{\today}

\maketitle

\section{Introduction}

Edge (and more generally, surface) states of quantum Hall (QH) and other topological states of matter provide a window to peek into the bulk topological properties of the system, via the principle of bulk/edge correspondence~\cite{Book}. More generally, an edge can be viewed as a special case of an interface between two different phases of matter, namely one of the two phases is a vacuum. Recently, interfaces between different QH phases have been attracting considerable attention~\cite{Grosfeld2009, Bais2009, Hughes2019, Regnault1, Regnault2, zhu20, Teo2020, Nielsen1, Nielsen2, wan16, Mross, Wang, Lian, simon20, Gefen2021, Heiblum2021}. While there is much similarity between edge and interface states, the latter is considerably richer because its physics depends on the topological properties of {\em both} phases. Nevertheless, the theoretical framework used thus far in their descriptions is the same.

The main purpose of the present paper is to address the important qualitative difference between edges and interfaces, and explore its consequences. For a QH edge, its location is usually determined by the confining potential that holds the electrons to form a QH liquid. This potential also determines the edge state spectrum, and other static and dynamical properties of the edge~\cite{wen-book}. The situation becomes very different for interfaces. To illustrate this point, consider magnetic domains of an Ising ferromagnet, in which the domain walls are the analogs of our interfaces. Due to the degeneracy of the two polarizations, the domain walls are free to move. While in reality the interfaces may be pinned by disorder or other extrinsic perturbations, there is no analog of the edge confining potential in the idealized limit. In such a limit, the interface becomes an extended string-like object free to move in the (2+1)-dimensional space-time~\cite{McGreevy}. Similarly, different QH phases may be realized in different regions of the sample. A prominent example of strong current interest is the fractional QH state at filling factor $\nu=5/2$ in GaAs heterostructures~\cite{Willett1987, Pan1999}, where the leading candidates, Pfaffian and anti-Pfaffian states, are exactly degenerate in the idealized limit (where particle-hole asymmetry and other extrinsic effects are absent). In this limit pinning effects of the domain walls disappear, and they are more appropriately described as string-like interfaces. On one hand, the low-energy physics of the system is still governed by area-preserving deformations of the QH liquid~\cite{Cappelli1993-1, Cappelli1993-2}, which are also known as edge waves~\cite{Stone1991}. On the other hand, the string-like nature of the interface significantly alters the spectra of the excitations and gives rise to fascinating physics. We believe the string-like nature of interfaces can play a fundamental role in understanding certain QH states. This viewpoint motivates our work.

In this paper, we study theoretically and numerically the low-energy excitation spectra of three different QH interfaces. In Sec.~\ref{sec:IQH}, we consider an interface between the Laughlin state at the filling factor $\nu=1/m$ and the vacuum. This simple setup allows us to revisit the physics of edge waves. We highlight the qualitative difference between low-energy excitations between free interfaces and pinned edges. In the special case of $m=1$, we provide a detailed analytic calculation of the effective string tension and compare it with our numerical results. Then, we discuss in Sec.~\ref{sec:MR-vac} the interface between the Pfaffian (also known as Moore-Read) state and the vacuum. Interestingly, we find that the energy scales of different types of excitations show an opposite hierarchy as compared to the low-energy spectrum of a pinned Pfaffian edge. Furthermore, we provide a theoretical analysis of the interface between Pfaffian and anti-Pfaffian states in Sec.~\ref{sec:Pf-Apf}. The understanding of this interface can be a crucial step to reveal the underlying nature of the $\nu=5/2$ fractional QH state. Specifically, we point out a more suitable description of the interface and clarify its difference from the usual description of pinned domains in existing literature. Finally, we summarize our work in Sec.~\ref{sec:summary} and outline briefly the possible analogies between QH interfaces and different concepts in string theory. Some technical details of calculation are given in the two appendices.

\section{Interface between Laughlin state and vacuum}
\label{sec:IQH}

\subsection{Low-energy effective theory}

\begin{figure} [htb]
\includegraphics[width=3.2in]{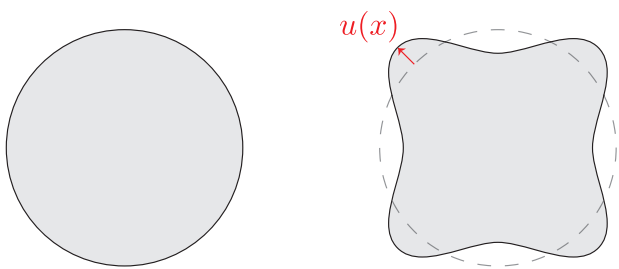}
\caption{Illustration of a quantum Hall droplet (left panel) in the ground state and in the presence of edge excitations, which can be understood as a distortion of its boundary (or edge wave) that preserves the area of the droplet (right panel).}
\label{fig:distortion}
\end{figure}

Let us start with the simple interface that separates a Laughlin state at filling factor $\nu=1/m$ and the vacuum~\cite{Laughlin}. When $m>1$ is odd (even), the Laughlin state describes fractional QH state of fermions (bosons) in the lowest Landau level. The Lagrangian describing the interface takes the form,
\begin{eqnarray} \label{eq:L}
L = L_0- H,
\end{eqnarray}
where the Hamiltonian $H$ is nonuniversal, and distinguishes between an edge and an interface. Meanwhile,
\begin{eqnarray}
L_0 = -\frac{m}{4\pi}\int{dx}(\partial_t\phi \partial_x\phi)
\label{eq:L0}
\end{eqnarray}
is a topological term~\cite{wen95}. Here, $x$ labels the coordinate along the edge or interface, and $\phi(x, t)$ is the chiral (right-moving) bosonic field. The physical meaning of $\phi(x, t)$ is determined by
\begin{eqnarray}
\rho(x)= \rho_0 u(x) = \frac{\partial_x\phi}{2\pi},
\end{eqnarray}
with $u(x)$ being the local distortion of the position of the edge/interface as illustrated in Fig.~\ref{fig:distortion}. The corresponding 1D electron density variation along the edge/interface is denoted as $\rho(x)$~\cite{Book}. We also use $\rho_0 = \nu/(2\pi\ell^2)$ to denote the 2D electron density at Landau level filling factor $\nu$, where $\ell=\sqrt{1/eB}$ is the magnetic length with $\hbar=c=1$.

We briefly review the edge excitations of the Laughlin state here. The $\nu=1/m$ QH liquid is formed by filling regions of low potential, which determines the shape and in particular the location of the edge~\cite{Book}. The energy cost of the distortion of edge comes from the potential energy cost. It takes the form
\begin{eqnarray}
H_e = \frac{\rho_0}{2}\int{dx}E(x)u^2(x),
\label{eq:Edge H}
\end{eqnarray}
where $E(x)$ is the (local) electric field at the edge that confines the electrons. Note that the electron charge is set to unity. For translational or rotational invariant edges, $E$ is a constant. Quantizing $H_e$ using Eq.~\eqref{eq:L0} leads to the familiar linearly dispersing chiral boson edge mode~\cite{wen95}.

Now, let us consider a physically very different case where the $\nu=1/m$ droplet is formed {\em spontaneously} due to {\em attractive} electron-electron interaction. Instead of filling some potential landscape, we now assume there is {\em no} external potential. In this case, the edge is more appropriately understood as the interface between the $\nu=1/m$ QH phase and the $\nu=0$ phase (or vacuum). Since this is a one-dimensional object, we will also refer to it as a string for reasons that will become clear soon. Different from edges, the energetics of the string-like interface is no longer determined by Eq.~\eqref{eq:Edge H}. Instead, it is set by the surface energy~\cite{GPS1994} which depends on the total length of the string~\cite{Zwiebach},
\begin{eqnarray} \label{eq:Interface H}
H_s = \frac{\sigma}{2}\int{dx}(\partial_x u)^2.
\end{eqnarray}
The symbol $\sigma$ labels the effective string tension or surface energy density. Quantizing $H_s$ (see Appendix~\ref{app:dispersion}) gives rise to a chiral bosonic interface mode with a cubic dispersion:
\begin{eqnarray}\label{eq:dispersion}
\omega(k)
= \frac{\sigma}{2\pi m\rho^2_0}k^3
= \sigma m(2\pi\ell)\ell^3 k^3.
\end{eqnarray}

\subsection{String tension for $\nu=1$ integer QH interface} \label{sec:tension}

The value of $\sigma$ can be determined from the finite-size correction in the ground state energy of the QH liquid. For a general Laughlin state at $\nu=1/m$, such an analytic calculation turns out to be challenging. The interface simplifies in to an integer QH interface in the special case of $m=1$. Here, we consider this simple (yet important) case and model the attractive electron-electron interaction by a single Haldane pseudopotential~\cite{Book},
\begin{eqnarray} \label{eq:pseudopotential}
V=V_1\sum_{i<j}\mathcal{P}_{i,j}^{l=1},
\end{eqnarray}
with $V_1<0$. The operator $\mathcal{P}_{i,j}^{l=1}$ projects the many-body wave function to the state, in which the two particles $i$ and $j$ have a relative angular momentum one~\cite{footnote-Laughlin}. It may be impractical to realize a single-order Haldane pseudopotential and attractive interaction between electrons in strongly-correlated electronic systems. At the same time, dilute ultracold atomic gases in rapidly rotating traps (in the lowest Landau level regime) provide a feasible platform to realize such an attractive interaction~\cite{Cooper-review}. In particular, the interaction in an ultracold one-component Fermi gas is dominated by the $p$-wave scattering, with all other components being strongly suppressed~\cite{Jin1999}. In the lowest Landau level regime, this interaction translates in to the $V_1$ pseudopotential~\cite{Regnault-fermion}. The $p$-wave interaction is usually very weak, but becomes more significant in dipolar gases~\cite{Baranov2002} and favors the formation of QH states~\cite{Baranov2005, Yi2011, Lewenstein2011, Yang2014}. Furthermore, the technique of Feshbach resonance enables one to tune the $p$-wave scattering length between atoms, and hence their interaction~\cite{Jin2003, Chin-RMP}. It also enables one to tune the scattering length to a negative value and realize an attractive interaction. Given the high tunability, it is hopeful that the $\nu=1$ and other interfaces can be realized in cold atom systems.

Suppose the QH droplet has $N$ electrons, then the unnormalized many-body wave function is~\cite{Book}
\begin{eqnarray} \label{eq:wave function}
\Psi(z_1,\cdots, z_N)
=\prod_{i<j}(z_i-z_j)
\exp{\left[-\sum_{i=1}^N\frac{|z_i|^2}{4\ell^2}\right]}.
\end{eqnarray}
In principle, one may calculate $\langle V\rangle$ for any value of $N$ from Eqs.~\eqref{eq:pseudopotential} and~\eqref{eq:wave function}, but the calculation is not straightforward. Using an alternative approach, we are able to evaluate $\langle V\rangle$ analytically (see Appendix~\ref{app:V-calculation}) and obtain
\begin{eqnarray}\label{eq:analytic-energy}
\langle V\rangle
=\frac{NV_1}{2}
\left[4-\frac{1}{4^{N-2}}\frac{(2N-1)!}{(N-1)!~ N!}
\right].
\end{eqnarray}

To study the finite-size correction in $\langle V\rangle$, we examine the asymptotic behavior of $\langle V\rangle$ as $N\rightarrow\infty$. By applying the Stirling's approximation, one obtains
\begin{eqnarray}\label{eq:asymptotic}
\langle V\rangle
=2NV_1-\frac{4V_1}{\sqrt{\pi}}\sqrt{N}+\mathcal{O}\left(\frac{1}{\sqrt{N}}\right).
\end{eqnarray}
The first term is the result in the thermodynamic limit, which is negative when $V_1<0$. It supports the spontaneous formation of the $\nu=1$ QH liquid due to the attractive electron-electron interaction. Meanwhile, the second term is positive and corresponds to the surface energy in the finite size system. The boundary effect is reflected in the
$\sqrt{N}$ dependence. For a disk-shaped QH droplet at $\nu=1$, its area is $2N\pi\ell^2$ and the corresponding circumference is $\sqrt{8N}\pi \ell$. Since the surface energy is proportional to the total length of the boundary due to the string-like behavior, the effective string tension $\sigma$ is
\begin{eqnarray}
\sigma\left(\sqrt{8N}\pi\ell\right)
=\frac{4|V_1|\sqrt{N}}{\sqrt{\pi}}.
\end{eqnarray}
Finally, we obtain
\begin{eqnarray} \label{eq:tension}
\sigma=\left(\frac{2}{\pi}\right)^{3/2}\frac{|V_1|}{2\ell}.
\end{eqnarray}
From the above result and Eq.~\eqref{eq:dispersion}, we determine the dispersion of the bosonic mode in the $\nu=1$ interface,
\begin{eqnarray}
\omega(k)=\sqrt{\frac{8}{\pi}}|V_1| \ell^3 k^3.
\end{eqnarray}
The same dispersion was derived from hydrodynamic theory~\cite{GPS1994} and the $W_{1+\infty}$ algebra~\cite{Cappelli1996}. From the dispersion, the result of $\sigma$ in Eq.~\eqref{eq:tension} was also deduced in Ref.~\cite{GPS1994}, but most of the details of the calculation were skipped there. Our analytic calculation fills the gap and confirms the result.
In ordinary edges the string tension is subdominant to the confining potential, which gives rise to a linear edge mode. Nonetheless, the string tension dominates the energetics of string-like interfaces, and leads to the interface mode with a cubic dispersion. This qualitative difference between interface and edge excitations may be probed by measuring the low-temperature specific heat $c(T)$. Notice that $T$ stands for temperature. Specifically, one expects $c(T)\sim T^{1/3}$ in an interface instead of a linear-$T$ dependence in an edge.

\subsection{Numerical studies for the $\nu=1$ integer QH interface}

\begin{figure} [htb]
\includegraphics[width=9cm]{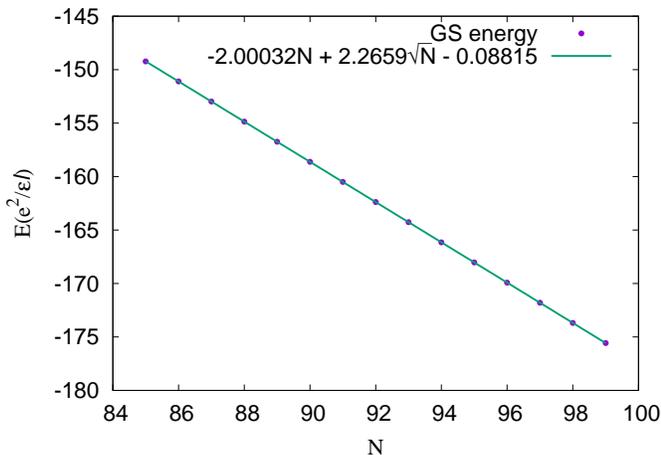}
\caption{Ground state energy of the $\nu=1$ integer QH interface as a function of electron number $N$. An attractive two-body interaction between electrons is included by setting $V_1=-1$ in Eq.~\eqref{eq:pseudopotential}. Here, we fit the numerical data with the equation, $\langle V \rangle = -2.0003N + 2.265\sqrt{N}-0.08815 $. The result agrees with Eq.~\eqref{eq:asymptotic} very well.}
\label{fig:IQHE-GS}
\end{figure}

Although previous numerical work has confirmed the existence of nonlinear excitations from edge deformations~\cite{Stone1992}, it remains desirable to verify the predicted value of the string tension. Motivated by this, we perform exact diagonalization to obtain the ground state energy and the low-energy excitation spectrum of the $\nu=1$ integer QH interface. In the simulation, we set the two-body interaction in the form of Eq.~\eqref{eq:pseudopotential} with $V_1=-1$. We consider the QH liquid in a disk-shaped geometry. The numerical results for the ground state energy when the system has  $N=85-99$ electrons are plotted in Fig.~\ref{fig:IQHE-GS}. First, we find that the numerical values agree with the analytic expression in Eq.~\eqref{eq:analytic-energy}. Also, we can fit the data very well by the expression $\langle V \rangle \approx -2.0003N + 2.266\sqrt{N}-0.08815$ as shown in the figure. Since both the coefficients of the $N$ and $\sqrt{N}$ terms are very close to $2V_1$ and $-4V_1/\sqrt{\pi}$, respectively, the asymptotic form of $\langle V\rangle$ in Eq.~\eqref{eq:asymptotic} is verified.

In addition, we check the cubic dispersion for the bosonic interface mode in Eq.~\eqref{eq:dispersion}. Since we are considering a disk-shaped geometry and applying the symmetric gauge, angular momentum $M$ is the good quantum number to label different electron orbitals. Hence, we need to relate the change in linear momentum $\delta k$ and the change in angular momentum $\Delta M$ of the system before we can discuss the excitation spectrum of the interface. For the $\nu=1$ integer QH liquid in the ground state, it has a radius $R_0 = \sqrt{2N}\ell$ and a total angular momentum $M_0 = N(N-1)/2$. For the sake of mimicking a smooth edge, the excitations in angular momentum subspace $M = M_0 + \Delta M$ need at least $N+\Delta M$ orbitals. As a result, the radius of the QH liquid is increased to $R = \sqrt{ 2 (N+\Delta M)}\ell$. The value of
$\delta k$ can be determined from $\delta M$ as
\begin{equation} \label{eq:delta k}
\delta k \simeq \frac{R-R_0}{\ell^2} \simeq \frac{\Delta M }{\sqrt{2 N}} \frac{1}{\ell}.
\end{equation}
Note that the relation only holds in the limit $\Delta M\ll N$.  In our numerical simulation, $\Delta M = 0 - 7$, and the number of electrons in the system ranged from $N = 40 - 99$. Hence, Eq.~\eqref{eq:delta k} is justified. We will use it to determine $\delta k$ in the following discussion.

\begin{figure} [htb]
\includegraphics[width=9cm]{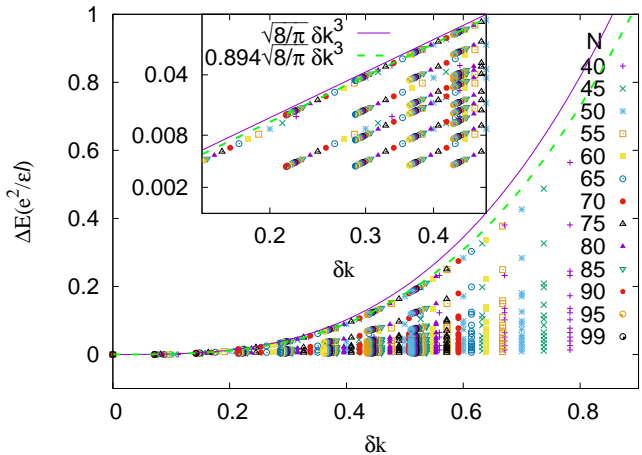}
\caption{The energy dispersion of the $\nu = 1$ integer QH interface with an attractive two-body interaction. Here, the system sizes range from $40$ to $99$ electrons. The change in energy $\Delta E$ (in units of $e^2/\epsilon \ell$) is plotted as a function of the additional momentum $\delta k $. The inset shows a log-log plot of the results in the small $\delta k$ region. We have collapsed our numerical results of the highest energy branch obtained from different system sizes into a single curve. Such curve is fit to the theoretical results, Eqs.~\eqref{eq:dispersion} and~\eqref{eq:tension}. For comparision, we also fit the same set of data to another curve with an adjustable coefficient $0.894$.}
\label{fig:IQHE-dispersion}
\end{figure}

By considering systems with $N = 40 - 99$ electrons, we plot the low-energy excitation spectrum of the interface as a function of $\delta k$ in Fig.~\ref{fig:IQHE-dispersion}. Similar to the lowest energy branch in FQH edges~\cite{Jain2014, jiang16}, here we find the {\em highest} excitation energies for each momentum of different systems collapse to a single curve; they correspond to the edge wave spectrum in our case. In the figure, $\Delta E  = E - E_0$ where $E_0$ given by Eq.~\eqref{eq:analytic-energy} stands for the ground state energy of the system. Here, $\Delta E$ is measured in units of $e^2/\epsilon\ell$, with $\epsilon$ being the dielectric constant of the system. From the figure, we find a zero energy mode in the lowest energy branch independent of the system size. This trivial mode comes from the center of mass motion. Importantly, we find that $\Delta E$ in the highest energy branch fits reasonably well to the dispersion relation $\omega(k)=(\sqrt{8/\pi})k^3$. The inset of Fig.~\ref{fig:IQHE-dispersion} highlights numerical results in the small $\delta k$ region.
Our numerical result verifies the cubic dispersion of the bosonic interface mode. It also verifies the effective string tension obtained from analytic calculation. Meanwhile, we found no signature of linearly dispersing bosonic modes, which clearly demonstrate the important qualitative difference between edge and interface excitations.

\section{Interface between the Pfaffian state and vacuum}
\label{sec:MR-vac}

\subsection{Theoretical considerations}

The previous section has set the stage for us to consider other (more) interesting interfaces. Here, we consider the interface between the Pfaffian (Moore-Read) state~\cite{moore91} and the vacuum. The Pfaffian state is well-known for hosting non-Abelian anyons which may be useful in topological quantum computation~\cite{TQC-RMP}. It has a ground state wave function~\cite{moore91},
\begin{align}
\nonumber
&\Psi_{\rm{MR}}(z_1, z_2, \cdots,z_N)
\\
=~ &\text{Pf}\left(\frac{1}{z_i-z_j}\right)
 \prod_{1\leq i<j\leq N}(z_i-z_j)^m
 \exp{\left[ -\sum_i^N \frac{\vert  z_i \vert^2}{4}\right]},
\end{align}
for a general filling factor $\nu=1/m$. Here, the notation Pf stands for the Pfaffian of the antisymmetric matrix with matrix elements $A_{ij}=1/(z_i-z_j)$. Physically, this Pfaffian factor comes from the Bardeen-Cooper-Schreiffer pairing of composite fermions in the chiral $p+ip$ channel~\cite{Read-Green}. Such a pairing leads to an additional Majorana fermion mode copropagating with the chiral bosonic mode along the interface~\cite{milovanovic96}. The Pfaffian interface is still described by the Lagrangian in Eq.~\eqref{eq:L} with $L_0$ given by~\cite{milovanovic96}:
\begin{eqnarray} \label{eq:L_MR}
L_0
=\int{dx}\left(-\frac{m}{4\pi}\partial_t\phi\partial_x\phi
-i\psi\partial_t\psi\right).
\end{eqnarray}
Here, $\psi(x,t)$ denotes the chiral Majorana fermion mode. Different from Laughlin states, the Pfaffian state provides a possible description of the QH state of fermions (bosons) when $m$ is even (odd). Besides the famous case $m=2$, previous work has suggested that the Pfaffian state of bosons at $\nu=1$ may be realized in cold atom systems~\cite{Gunn2000, Cooper2001, Regnault2003, Jain2005, Regnault2007, Viefers2008}. This state is the exact zero-energy ground state of the three-body interaction,
\begin{eqnarray}
H_{3B}=V_{3B}\sum_{i<j<k}\delta(\mathbf{r}_i-\mathbf{r}_j)\delta(\mathbf{r}_j-\mathbf{r}_k).
\end{eqnarray}
In a real sample, $H_{3B}$ can be a consequence of the Landau-level mixing effect.

In addition to $H_{3B}$, we assume there is an attractive two-body interaction between the bosons, so that a Pfaffian droplet can be formed spontaneously. As a simple example, we consider the Haldane pseudopotential,
\begin{eqnarray} \label{eq:V-MR}
V=V_0\sum_{i<j}\mathcal{P}_{i,j}^{l=0},
\end{eqnarray}
with $V_0<0$. It is expected that the bosonic mode $\phi$ still satisfies a cubic dispersion since its excitations still originate from deforming the interface. Although we have not obtained an analytic result of the string tension for the Pfaffian interface, it should be different from Eq.~\eqref{eq:tension}. For the additional Majorana fermion mode, it gives rise to fermionic excitations of the interface. Importantly, such excitations \textit{do not} come from the deformation of interface. Instead, they come from the breaking of Cooper pairs of composite fermions~\cite{milovanovic96}. It is reasonable to guess the fermionic and bosonic interface modes have different dispersions and energetics. As we are going to demonstrate, such an intriguing feature is indeed validated by our numerical results. Note that the above feature should also exist in a Pfaffian interface with $m>1$.

\subsection{Numerical studies of the $\nu=1$ Pfaffian interface}

\begin{table*} [htb]
\begin{spacing}{1.2}
\begin{center}
\begin{tabular*}{\hsize}{@{}@{\extracolsep{\fill}}ccc@{}}
\hline \hline
 $\Delta M$
&  bosonic mode configurations
& counting number \\
\hline
$0$ & 0  & 1\\
$1$ & 1  & 1\\
$2$ & 2, 1+1  & 2\\
$3$ & 3, 2+1, 1+1+1 & 3\\
$4$ & 4, 3+1, 2+2, 2+1+1, 1+1+1+1 &  5\\
$5$ & 5, 4+1, 3+2, 3+1+1, 2+2+1, 2+1+1+1, 1+1+1+1+1 &  7 \\
\end{tabular*}
\begin{tabular*}{\hsize}{@{}@{\extracolsep{\fill}}ccc@{}}
\hline
 $\Delta M$
& fermionic mode configurations
& counting number \\
\hline
$0$ & null & null\\
$1$ & null  & null\\
$2$ & $\frac{1}{2}+\frac{3}{2}$  & 1\\
$3$ & $\frac{1}{2}+\frac{5}{2}$ & 1\\
$4$ & $\frac{1}{2}+\frac{7}{2}, \frac{3}{2}+\frac{5}{2}$ &  2\\
$5$ & $\frac{1}{2}+\frac{9}{2}, \frac{3}{2}+\frac{7}{2}$ &  2 \\
\end{tabular*}
\begin{tabular*}{\hsize}{@{}@{\extracolsep{\fill}}ccc@{}}
\hline
 $\Delta M$
&  convoluted mode configurations
& counting number \\
\hline
$0$ & null & null\\
$1$ & null  & null\\
$2$ & null  & null\\
$3$ & $1+\frac{1}{2}+\frac{3}{2}$ & 1\\
$4$ & $2+\frac{1}{2}+\frac{3}{2}, 1+1+\frac{1}{2}+\frac{3}{2} , 1+\frac{1}{2}+\frac{5}{2}$ &  3\\
$5$ & $3+\frac{1}{2}+\frac{3}{2}, 2+1+\frac{1}{2}+\frac{3}{2} , 1+1+1+\frac{1}{2}+\frac{3}{2}, 2+\frac{1}{2}+\frac{5}{2}, 1+1+\frac{1}{2}+\frac{5}{2}, 1+\frac{1}{2}+\frac{7}{2},1+\frac{3}{2}+\frac{5}{2}$ &  7 \\
\hline
\end{tabular*}
\end{center}
\caption{Possible configurations of $\Delta M$ ranging from 0 to 5. Here, we classify the configurations as pure bosonic mode $(n_f=0)$, pure fermionic mode $(n_b=0)$, and the convoluted mode $(n_b\neq 0, n_f\neq 0)$. The corresponding numbers of different configurations for each $\Delta M$ (counting number) are listed in the rightmost column.}
\label{tb: counting-number}
\end{spacing}
\end{table*}

We assume each low-energy excitation of the interface can be described by a collection of bosonic and fermionic excitations, and their convolution. For the bosonic excitation with an integer angular momentum $l_b$, its energy is $\epsilon_b(l_b)$. Such a bosonic mode can be occupied by ${n_b(l_b)}$ bosons. Similarly, we denote the energy of a fermionic excitation as
$\epsilon_f(l_f)$, which has a half-integer angular momentum $l_f$. Different from ${n_b(l_b)}$ which can be any non-negative integer, each fermionic mode can only accommodate $n_f(l_f) = 0, 1$ fermion due to the Pauli exclusion principle. Note that the total fermion occupation number $\sum_{l_f}n_f(l_f)$ must be even. It is because an even number of Majorana fermions are produced by breaking Cooper pairs of composite fermions. Based on the above description, the energy of a low-energy excitation of the Pfaffian interface is given by
\begin{eqnarray} \label{eq:Delta-E}
\Delta E = \sum_{l_b} n_b(l_b) \epsilon_b(l_b) + \sum_{l_f} n_f(l_f)\epsilon_f(l_f).
\end{eqnarray}
$\Delta E=E-E_0$ carries the same physical meaning as in the previous case. Furthermore, the change in total angular momentum of the system from $M_0$ is
\begin{eqnarray}
\Delta M = \sum_{l_b} n_b(l_b) l_b + \sum_{l_f} n_f(l_f) l_f.
\end{eqnarray}
Here, $M_0=N(N-2)/2$ is the total angular momentum of the ground state of the Pfaffian droplet with $N$ bosons. Since our simulation is limited to $\Delta M = 0 - 5$, low-energy excitations obtained from the simulation can be associated to any one of the possible configurations listed in Table.~\ref{tb: counting-number}.

We perform exact diagonalization to study numerically the excitation spectrum of the Pfaffian interface. In the simulation, we set $H_{3B} = 2\pi^2 \ell^4 \sum_{i<j<k} \delta(\textbf{r}_i-\textbf{r}_j) \delta(\textbf{r}_j-\textbf{r}_k)$. The prefactor $2\pi^2 \ell^4$ is chosen to ensure the three-body pseudopotential is properly normalized. For the attractive two-body interaction, we set $V_0=-1$ in Eq.~\eqref{eq:V-MR}. The mixed Hamiltonian that we diagonalize takes the form,
\begin{equation}\label{eq:mixedH}
H = H_{2B}(V_0) + \lambda H_{3B}.
\end{equation}
The parameter $\lambda$ is set to a very large number (essentially infinite) so that the Pfaffian state becomes energetically favorable.

\begin{figure} [htb]
\includegraphics[width=9cm]{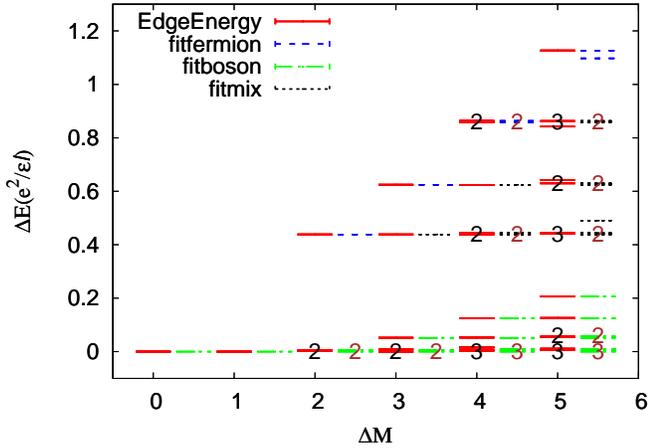}
\caption{Low-energy excitations $\Delta E(\Delta M)$ from exact diagonalization of the mixed Hamiltonian in Eq.~\eqref{eq:mixedH}. Here, we included $N=10$ bosons and $14$ orbitals in the simulation. Fitting the numerical results with Eq.~\eqref{eq:Delta-E} classifies different excited states of the interface as having only bosonic excitations, or only fermionic excitations, or a mixture of both excitations. These are labeled by blue dashed bars, green dot-dash bars, and black dotted bars, respectively. For a better visualization, we transverse all the fitted bars to the right by $\Delta M=0.5$. The numbers on thicker bars represent the degeneracy of the degenerate states and the unlabelled ones are nondegenerate states.}
\label{fig:MRN10-spectrum}
\end{figure}

Let us first focus on the results from the system with $N=10$ bosons in a total number of $14$ possible orbitals. After diagonalizing the mixed Hamiltonian $H$, we obtained the energy spectrum as shown in Fig.~\ref{fig:MRN10-spectrum}. Specifically, the numerical results of $\Delta E$ for different excited states are shown as red solid lines in the figure. When we count the total number of excitations at each value of $\Delta M$, we obtain the sequence $1, 1, 3, 5, 10, 16$. This result agrees with the total counting numbers in Table~\ref{tb: counting-number} by including all three types of excitations. It is desirable to analyze the spectrum in more detail and classify each excitation as pure bosonic, or pure fermionic, or the convoluted type. We follow the analysis in Ref.~\cite{wan08} closely and apply Eq.~\eqref{eq:Delta-E} to fit our numerical results of $\Delta E$. The results from the analysis are also illustrated in Fig.~\ref{fig:MRN10-spectrum}, next to the numerical results. Similar to the case of Pfaffian edge, we find that the energy spectrum of the Pfaffian interface also separates into two sections. This separation occurs at around $\Delta E=0.4 e^2/\epsilon l$. For the low-energy section with $\Delta E<0.4 e^2/\epsilon l$, those excitations are identified as pure bosonic. In other words, only the bosonic interface mode $\phi$ is excited due to the deformation of the interface. One may also count the numbers of such low-energy states for each value of $\Delta M$. This gives $1, 1, 2, 3 ,5, 7$, which match precisely the counting numbers for pure bosonic modes in Table.~\ref{tb: counting-number}. Since there is no confining potential of the QH liquid, it costs zero energy to excite the center of mass mode. Thus, we have $\epsilon_b(l_b=1) = 0$. For the high-energy section with $\Delta E> 0.4 e^2/\epsilon\ell$, the excitations are either pure fermionic or the convoluted type. More specifically, we discover the pure fermionic excitations have the highest energy scale, whereas the convoluted type of excitations have their energy scale in between the pure fermionic and the pure bosonic excitations. It is very striking to point out that this hierarchy of energy scales is opposite to the one in the Pfaffian edge (for fermions in the half-filled Landau level) pinned by a confining potential~\cite{wan08, Jain2014, jiang16}. There, the highest energy scale is occupied by pure bosonic modes, whereas the pure fermionic modes have the lowest energy scale. We will get back to this point later.

\begin{figure} [htb]
\includegraphics[width=9cm]{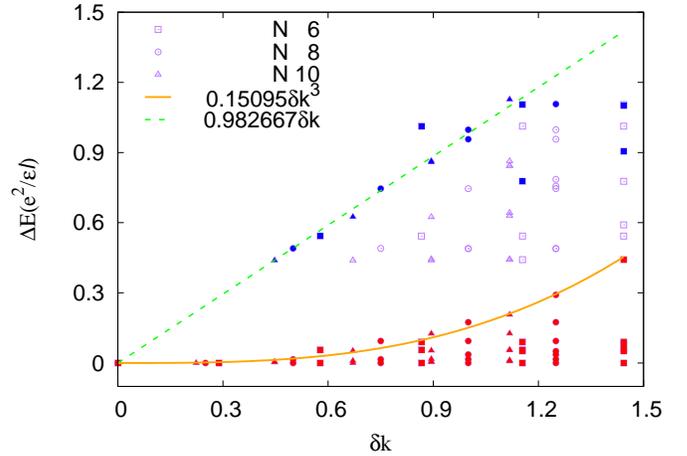}
\caption{The energy dispersion of the $\nu = 1$ Pfaffian interface with an attractive two-body interaction and a large repulsive three-body interaction. Numerical results for system sizes $N = 6, 8, 10$ are marked by squares, circles and triangles, respectively. Different energy states are classified as having pure fermionic (blue filled points), pure bosonic (red filled points), and convoluted (purple open points) excitations. Data points in the lower energy branches are recognized as pure bosonic mode with the fitting function $\omega(\delta k) = 0.150953 \cdot {\delta k}^3$. The top branches of the higher energy section are fermionic mode with a linear fitting $\omega(\delta k)= 0.98266 \cdot \delta k$. Note that the linear momentum is determined from $\delta k = \Delta M / \sqrt{2N}\ell$. See the main text for more details.}
\label{fig:MR-dispersion}
\end{figure}

To have a better understanding of the energetics, we investigate the energy dispersions of the low-energy excitations in the $\nu=1$ Pfaffian interface. The mixed Hamiltonian in Eq.~\eqref{eq:mixedH} for systems with $6$, $8$, and $10$ bosons were diagonalized. Using Eq.~\eqref{eq:delta k}, we convert $\Delta M$ into $\delta k$, and plot the numerical results of $\Delta E(\delta k)$ with different system sizes in Fig.~\ref{fig:MR-dispersion}. A numerical fitting reveals the highest branch of the pure bosonic excitations (originating from the bosonic interface mode) still satisfies a cubic dispersion. This is also illustrated in the same figure. Consistent with our theoretical prediction, the dispersion is not identical to the one in $\nu=1$ integer QH interface of electrons. In the present case, we obtain the fitting curve as $\omega_b(\delta k) \approx 0.15 {\delta k}^3$. This result allows us to predict the effective string tension of the $\nu=1$ Pfaffian interface:
\begin{eqnarray}
\sigma_{\rm MR}
\approx \frac{0.15}{2\pi} \frac{|V_0|}{\ell},
\end{eqnarray}
where the linear dependence in $V_0/\ell$ is deduced from a simple dimensional analysis. In contrast to the bosonic excitations, our numerical data show that pure fermionic excitations (originating from the Majorana fermion mode along the interface) satisfy a linear dispersion. We obtain the corresponding best fit line as $\omega_f (\delta k) \approx 0.98 \delta k$. A simple dimensional analysis suggests that the speed of the fermion mode $v_f\sim |V_0|\ell$ (with $\hbar=1$). However, a detailed theory to explain the energetics is still lacking, going back to the hierarchy of energy scales in different types of excitations. Since we focus on the low-energy excitations at small $\delta k$, the cubic dispersion of the bosonic interface mode strongly reduces its energy compared to the fermionic mode. It is not surprising that a crossover between these two energy scales may occur at a larger value of $\delta k$, which has not been reached by our simulation.

\section{Interface between Pfaffian and anti-Pfaffian states}
\label{sec:Pf-Apf}

In this section, we provide a theoretical analysis of the interface between the Pfaffian and the anti-Pfaffian (APf) states~\cite{wan16, BMF15}. The APf state is the particle-hole conjugate of the Pfaffian state~\cite{lee07, levin07}, which was introduced as another candidate to describe the fractional QH state at $\nu=5/2$ in GaAs heterostructures~\cite{Willett1987, Pan1999}. However, neither of them can explain all existing experimental results in a natural way. Motivated by this, Wan and Yang introduced an intermediate state, which consists of alternating Pfaffian and APf stripes in the bulk of a realistic sample~\cite{wan16}. This idea was later modified to explain the ``unexpected" result from thermal Hall conductance experiment~\cite{Banerjee2018}. In particular, mesoscopic puddles of Pfaffian and APf domains can form in the bulk of a realistic sample due to the presence of disorder~\cite{Mross, Wang, Lian}. This proposal has been substantiated by a recent work, which pointed out that the formation of Pfaffian-APf domain walls may be energetically favorable in realistic samples~\cite{zhu20}. In the idealized limit of vanishing Landau-level mixing effect and no disorder-mediated density modulation, the APf and Pfaffian states are exactly degenerate~\cite{lee07, levin07}. These symmetry-breaking effects must exist in realistic samples, but they can be rather small. The domain walls are still free to propagate rather than being strongly pinned in the sample. In this scenario, it is more appropriate to view the domain wall as a string-like interface.

\begin{figure} [htb]
\includegraphics[width=3.2in]{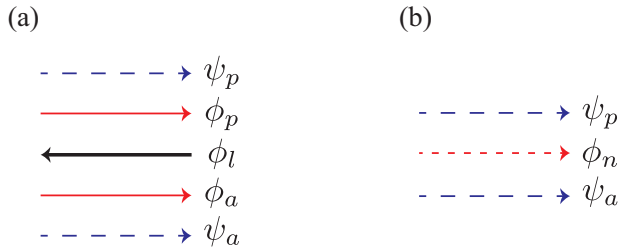}
\caption{(a) The original modes in the Pfaffian and anti-Pfaffian edges. (b) The resulting modes in the Pfaffian-anti-Pfaffian interface due to a relevant electron-pair tunneling across the edges. See the main text for the details of each mode.}
\label{fig:Pf-APf}
\end{figure}

Here, we follow Refs.~\cite{BMF15, wan16} and give a brief review of the Pfaffian-APf interface. At the interface, the Pfaffian and APf edges have opposite chiralities, which are described by the Lagrangian~\cite{milovanovic96, lee07, levin07},
\begin{align}
\nonumber
L
=~&\int{dx} \left(-\frac{2}{4\pi}\partial_t\phi_p\partial_x\phi_p
-i\psi_p\partial_t\psi_p\right)
\\ \nonumber
&+\int{dx} \left(\frac{1}{4\pi}\partial_t\phi_l\partial_x\phi_l
-\frac{2}{4\pi}\partial_t\phi_a\partial_x\phi_a - i\psi_a\partial_t\psi_a\right)
\\
&-H.
\end{align}
The first line describes the Pfaffian edge which has a bosonic mode $\phi_p$ and a Majorana fermion mode $\psi_p$. The second line describes the APf edge that contains two counterpropagating bosonic modes $\phi_l$ and $\phi_a$, and a Majorana fermion mode
$\psi_a$. Finally, the Hamiltonian $H$ describes the interaction between different edge modes, which eventually determines the structure of the interface. By including electron-pair tunneling between the edges, the two charge modes $\phi_l$ and $\phi_r=\phi_p+\phi_a$ are gapped. In the presence of strong Coulomb interaction, the tunneling process is relevant in the renormalization group sense. As a result, the remaining gapless modes at the interface are the \textit{neutral} bosonic mode $\phi_n=\phi_p-\phi_a$, and the pair of Majorana fermion modes, $\psi_p$ and $\psi_a$. The topological term describing the interface modes is~\cite{BMF15, wan16},
\begin{eqnarray} \label{eq:L-Pf-APf}
L_0
=\int{dx}\left(-\frac{1}{4\pi}\partial_t\phi_n\partial_x\phi_n
-i\psi_p\partial_t\psi_p
-i\psi_a\partial_t\psi_a
\right).
\end{eqnarray}
In Fig.~\ref{fig:Pf-APf}, different modes in the original Pfaffian and APf edges, and the resulting interface are illustrated.

Now, it is important for us to emphasize the difference between Eq.~\eqref{eq:L-Pf-APf} and another common description in the existing literature. It has been pointed out that there are four copropagating Majorana fermions at the Pfaffian-APf domain wall. Depending on how they are localized, different phases can be realized in the bulk and at the edge of the system~\cite{Mross, Wang, Lian}. This four-Majorana picture is a natural description for pinned domains by disorder, which are usually assumed in the existing literature. Naively, one may recover this description by fermionizing $\phi_n$ into two Majorana fermions that have the same chirality as $\psi_p$ and $\psi_a$. However, our following discussion suggests that Eq.~\eqref{eq:L-Pf-APf} turns out to be a more natural description of the Pfaffian-APf interface.

First, we show that $\phi_n$ satisfies a cubic dispersion due to the distortion of the string-like interface. Since $\phi_n=\phi_p-\phi_a$, we deduce its energy dispersion starting from
\begin{align}
\label{eq:u-p}
\rho_{1/2} u_p(x)=\frac{u_p(x)}{4\pi\ell^2}=\frac{\partial_x\phi_p}{2\pi},
\\
\label{eq:u-a}
\rho_{1/2} u_a(x)=\frac{u_a(x)}{4\pi\ell^2}=\frac{\partial_x\phi_a}{2\pi}.
\end{align}
Here, $\rho_{1/2}=(1/2)/(2\pi\ell^2)$ is the electron density in a half-filled Landau level. The symbols $u_p(x)$ and $u_a(x)$ denote the local distortions of the positions of the original
$\nu=1/2$ edges in the Pfaffian and anti-Pfaffian edges, respectively. The surface energy of the interface is
\begin{eqnarray}
H_s=\frac{\sigma}{2}\int \left[\partial_x u_p(x)-\partial_x u_a(x)\right]^2~dx.
\end{eqnarray}
Combining Eqs.~\eqref{eq:u-p} and~\eqref{eq:u-a}, we can write $H_s$ as
\begin{align}
\nonumber
H_s=~&\frac{\sigma}{2}\left(\frac{1}{2\pi\rho_{1/2}}\right)^2
\int{dx} \left(\partial_x^2\phi_n\right)^2
\\
=~&H_0+\frac{\sigma}{\left(2\pi\rho_{1/2}\right)^2}
\sum_{k>0}k^4\phi_{n,k}\phi_{n,k}^\dagger.
\end{align}
In the second equality, we have expanded $\phi_n(x)$ in the plane-wave basis with Fourier modes $\phi_{n,k}$. Using the commutation relation
$[\phi_{n,k}, \phi_{n,k'}^\dagger]=(2\pi/k')\delta_{k,k'}$, we obtain the energy dispersion for the interface mode $\phi_n$:
\begin{eqnarray}
\omega_n(k)
=\frac{\sigma}{2\pi\rho_{1/2}^2}k^3
=\sigma (8\pi \ell) \ell^3 k^3.
\end{eqnarray}
Due to the above cubic dispersion relation, it is unnatural to fermionize $\phi_n$ into two Majorana fermions. Thus, the four-fermion picture is not the most suitable description of a string-like interface. Based on a density matrix renormalization group calculation, the domain wall tension of the interface was estimated as $\sigma\approx (2.2\times 10^{-3}) e^2/\epsilon\ell^2$ in Ref.~\cite{zhu20}. This value is consistent with Ref.~\cite{simon20}. Hence, we predict the dispersion for $\phi_n$ as $\omega_n(k)\approx (5.5\times 10^{-2})(e^2/\epsilon\ell)\ell^3k^3$.

Next, we expect both $\psi_p$ and $\psi_a$ would satisfy linear dispersions. In a general scenario, we do not see any natural reason for having a symmetry between $\psi_p$ and $\psi_a$. Therefore, they have different speeds, and their linear energy dispersions are not identical. In addition, we believe the excitations with the lowest energy scale are still pure bosonic excitations. Since the interface has three different gapless modes, it is not surprising that its spectrum can be quite complicated. Nevertheless, we believe all our predictions here can be checked in future numerical work. In our opinion, such a checking will be a crucial step to understand the nature of the Pfaffian-APf domain wall, and hence the underlying nature of the $\nu=5/2$ fractional QH state in a real sample.

\section{Summary and Outlook: More general interfaces and edge as a D-brane}
\label{sec:summary}

To summarize our work, we have studied three different QH interfaces to highlight the qualitative differences between interfaces and edges. In ordinary QH edges, they are pinned by confining potentials. On the other hand, the interfaces are string-like and free to propagate in the system. Such an interface forms when the QH liquid is held together by attractive interaction between particles. Despite the fact that both the low-energy excitations of pinned edges and interfaces are edge waves (area-preserving deformation of the QH liquid), we showed that the bosonic interface mode satisfies a cubic dispersion instead of the ubiquitous linear dispersion in edge excitations. Our work shows that these excitations dominate the low-energy physics of interfaces. This qualitative feature may be probed in a low-temperature specific heat measurement, which is expected to show $c(T)\sim T^{1/3}$.

In the simple case with a $\nu=1$ integer QH state held by an attractive Haldane pseudopotential $V_1<0$, we have derived analytically the effective string tension of the interface. Our numerical results confirm the theoretically predicted value $\sigma=(2/\pi)^{3/2}|V_1|/2\ell$. The cubic dispersion of the bosonic interface mode has also been verified by our numerical results. Furthermore, we have studied numerically the low-energy spectrum of the Pfaffian (Moore-Read) interface for bosons at $\nu=1$, which is formed by including the two-body Haldane pseudopotential $V_0<0$ and a large repulsive three-body interaction. Based on the numerical results, we made a prediction to the string tension of the interface as $(0.15/2\pi)|V_0|/\ell$. The presence of the fermionic interface mode complicates the spectrum, but also makes it more interesting. By identifying the nature of different excited states, we discovered the excitations with the lowest energy scale correspond to pure bosonic excitations, whereas the high energy scale is occupied by pure fermionic excitations whose dispersion remains linear. Such a hierarchy of energy scales is opposite to the one in a Pfaffian edge~\cite{wan08, Jain2014}.

Simple as they may be, much of our considerations also apply to interfaces between different QH liquids, as long as they are {\em not} pinned. This principle motivated us to consider the interface between Pfaffian and anti-Pfaffian (APf) states. Recent studies assumed the interfaces are pinned by disorder potential~\cite{Mross, Wang, Lian}, except for Ref.~\cite{wan16} which suggested the spontaneous formation of alternating Pfaffian and anti-Pfaffian stripes, and corresponding interfaces. When the interaction between different interfaces is ignored, the Pfaffian-APf interface should demonstrate similar features of the two simpler interfaces discussed above. In particular, we argue that the neutral bosonic mode satisfies a cubic dispersion, whereas the two Majorana fermion modes satisfy different linear dispersions. Based on the recent results of the domain wall tension~\cite{simon20, zhu20}, we have predicted the cubic dispersion as $\omega_n(k)=\alpha k^3$ with $\alpha\approx (5.5\times 10^{-2})(e^2\ell^2/\epsilon)$. Such a nonlinear dispersion suggests that one should not fermionize the neutral bosonic mode into two Majorana fermions. Thus, the interface modes should not be described as four copropagating Majorana fermions. In fact, the picture of having one bosonic and two Majorana fermion modes has been proposed in Refs.~\cite{BMF15, wan16}. We hope our predictions can be verified in future numerical simulations.

\begin{figure} [htb]
\includegraphics[width=1.75in]{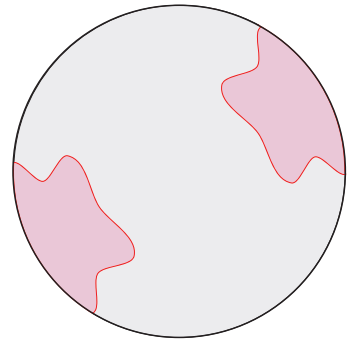}
\caption{Illustration of open strings (red lines) attached to a quantum Hall edge (black circle) determined by confining potential. Regions shaded with different colors are in different quantum Hall phases. The edge plays a role very similar to a D-brane in string theory.}
\label{fig:brane}
\end{figure}

Finally, it is tempting to make analogy between quantum Hall interfaces and different concepts in string theory. From the nature of the interface modes, the Laughlin and Pfaffian interfaces behave like bosonic string and superstring~\cite{SUSY-QH}, respectively. In this paper, the interfaces we have considered can only form closed strings. In more generic cases where an edge (determined by confining potential) is also present, more features can be observed. For example, the presence of the edge allows one to include open strings by anchoring the interface on the edge as demonstrated in Fig.~\ref{fig:brane}. In this sense, the edge plays a role like a D-brane~\cite{Dbrane} (more specifically a D1-brane since the edge is one-dimensional) in string theory. It is known that the phase of the D-brane is sensitive to the state of the open strings~\cite{Dbrane}. In the quantum Hall analogy, there is much interest and debate on the relation between the interfaces and edge state in the $\nu=$5/2 fractional QH state~\cite{wan16,Mross,Wang,Lian,simon20,zhu20}. It will be very interesting to pursue the analogy deeper, and seek the possibility of using quantum Hall systems to study different abstract concepts in string theory. We believe the present paper provides an interesting perspective on this fascinating physics.

\section*{Acknowledgments}

The work of K. K. W. M., R. W. and K. Y. was performed at the National High Magnetic Field Laboratory, which is supported by National Science Foundation Cooperative Agreement No. DMR-1644779, and the State of Florida. Z.-X. H. was supported by National Natural Science Foundation of China Grant No. 11974064 and the Fundamental Research Funds for the Central Universities under Grant No. 2020CDJQY-Z003. H. W. was supported by the Guangdong Provincial Key Laboratory (Grant No. 2019B121203002). R. W. was supported by DOE grant No. \protect{DE-SC0002140}. K. Y. was supported by the National Science Foundation Grant No. DMR-1932796. 

Q.L. and K.K.W.M. contributed equally to this work.

\appendix

\section{Derivation of the cubic dispersion for bosonic interface mode}
\label{app:dispersion}

In this appendix, we derive the cubic dispersion of the bosonic interface mode. We start by writing $\phi(x)$ in the plane-wave basis:
\begin{eqnarray} \label{eq:plane-wave}
\phi(x)
=\sum_k \frac{1}{\sqrt{L}}e^{ikx}\phi_k.
\end{eqnarray}
Here, $L$ denotes the length of the interface. Since $\phi(x)$ is a real field, the Fourier mode $\phi_k$ satisfies $\phi_{-k}=\phi_k^\dagger$. A direct substitution of Eq.~\eqref{eq:plane-wave} in Eq.~\eqref{eq:Interface H} gives
\begin{eqnarray}
H_s=H_0+\frac{\sigma}{(2\pi\rho_0)^2}\sum_{k>0}k^4\phi_k^\dagger \phi_k,
\end{eqnarray}
where $H_0$ is a constant. For simplicity, we set $H_0=0$ in the following discussion. Recall that any excited state with a linear momentum $k$ is generated by $\phi_k^\dagger|\rm{vac}\rangle$, where the vacuum state $|\rm{vac}\rangle$ satisfies $\phi_k|\rm{vac}\rangle=0$ for all $k>0$. Acting $H_s$ on the excited state, one gets
\begin{align} \label{eq:result}
\nonumber
&\frac{\sigma}{(2\pi\rho_0)^2}\sum_{q>0}q^4\phi_q^\dagger \phi_q
\left(\phi_k^\dagger|\rm{vac}\rangle\right)
\\ \nonumber
=~&\frac{\sigma}{(2\pi\rho_0)^2}\sum_{q>0}q^4\phi_q^\dagger
\left[
\frac{2\pi}{mk}\delta_{q,k}+\phi_k^\dagger\phi_q\right]
|\rm{vac}\rangle
\\
=~&\frac{\sigma}{2\pi m\rho_0^2}k^3\left(\phi_k^\dagger |\rm{vac}\rangle\right).
\end{align}
Note that we have used the commutation relation,
$[\phi_k,\phi_{k'}^\dagger]=(2\pi/mk')\delta_{k,k'}$~\cite{wen-book} in the above calculation. From Eq.~\eqref{eq:result}, we derive the cubic dispersion in Eq.~\eqref{eq:dispersion} in the main text.

\section{Calculation of $\langle V\rangle$ for the $\nu=1$ integer QH interface}
\label{app:V-calculation}

Instead of using Eqs.~\eqref{eq:pseudopotential} and~\eqref{eq:wave function}, we evaluate the correlation function,
\begin{eqnarray}
g(z_1,z_2)
=\frac{N(N-1)}{\rho_0^2}
\frac{\int~ dz_3\cdots dz_N~ |\Psi|^2}
{\int~ dz_1\cdots d z_N~ |\Psi|^2}.
\end{eqnarray}
For $N$ being finite, it is remarkable that $g(z_1,z_2)$ still has a closed form:
\begin{widetext}
\begin{eqnarray} \label{eq:finite-N}
g(z_1,z_2)
=\frac{\Gamma(N, |z_1|^2/2\ell^2)}{(N-1)!}
\frac{\Gamma(N, |z_2|^2/2\ell^2)}{(N-1)!}
-\frac{\Gamma(N, \bar{z}_1 z_2/2\ell^2)}{(N-1)!}
\frac{\Gamma(N, z_1\bar{z}_2 /2\ell^2)}{(N-1)!}
e^{-|z_1-z_2|^2/2\ell^2}.
\end{eqnarray}
\end{widetext}
Here, $\Gamma(N,x)$ denotes the upper incomplete Gamma function. Having $g(z_1,z_2)$ in hand, a more useful expression for the interaction in Eq.~\eqref{eq:pseudopotential} is the real space representation formulated by Trugman and Kivelson~\cite{Trugman1985}:
\begin{eqnarray} \label{eq:TK-form}
V_{\rm TK}
=(4\pi\ell^2)V_1L_1(-\ell^2\nabla^2_r)\delta^2(r).
\end{eqnarray}
The function $L_1(x)$ stands for the first-order Laguerre polynomial, whereas $r=|z_2-z_1|$ denotes the separation between the two particles. When $z_1=z_2$, $g(z_1,z_2)$ vanishes. Thus, one has
\begin{eqnarray}
\langle V\rangle
=\frac{\rho_0^2}{2}\int  (4\pi\ell^4)V_1\delta^2(r)
\nabla^2_r g(z_1,z_2)~dz_1 dz_2.
\end{eqnarray}

From the above discussion, it suffices to fix $z_1$ and expand $g(z_1,z_2)$ up to $r^2$ to determine $\langle V\rangle$. To be specific, only the term proportional to $r^2$ matters. We obtain the corresponding term
\begin{eqnarray}
g(x,r)
=\frac{r^2}{2\ell^2}~f(N,x) + \text{other terms}.
\end{eqnarray}
In the above equation, we have introduced the dimensionless variable $x=|z_1|^2/2\ell^2$. Furthermore, the dimensionless function $f(N,x)$ is defined as
\begin{widetext}
\begin{eqnarray}  \label{eq:expand-r2}
f(N,x)
=\frac{[\Gamma\left(N,x\right)]^2-x^{2N-1}e^{-2x}
+\left(x^N-Nx^{N-1}\right)\Gamma\left(N, x\right)e^{-x}}
{[(N-1)!]^2}.
\end{eqnarray}
\end{widetext}
 Then, we arrive at
\begin{eqnarray}
\langle V\rangle
=2V_1\int_0^\infty~ f(N,x)~dx.
\end{eqnarray}
The above integral can be computed analytically, and leads to Eq.~\eqref{eq:analytic-energy} in the main text.


\begin{thebibliography}{99}
\bibitem{footnote-Laughlin}
For $m>1$, the Laughlin wave function is a zero-energy ground state of the two-body Haldane pseudopotentials $V_{m'}$ with $m'<m$. Hence, a spontaneous formation of the
$\nu=1/m$ Laughlin droplet would require at least one of the components $V_{m'}<0$ for $m'\geq m$. The value of $\langle V\rangle$ may be obtained numerically. In the main text, we consider analytically the particular case $m=m'=1$.
\end{thebibliography}
\end{document}